# Observation of the nonlinear anomalous Hall effect in 2D WTe$_2$

Kaifei Kang[1], Tingxin Li[1], Egon Sohn[2,3], Jie Shan[1,2,4*], and Kin Fai Mak[1,2,4*]
[1] School of Applied and Engineering Physics, Cornell University, Ithaca, NY, USA
[2] Laboratory of Atomic and Solid State Physics, Cornell University, Ithaca, NY, USA
[3] Department of Physics, Penn State University, University Park, PA, USA
[4] Kavli Institute at Cornell for Nanoscale Science, Ithaca, NY, USA

*E-mails: jie.shan@cornell.edu; kinfai.mak@cornell.edu.
These authors contributed equally: Kaifei Kang, Tingxin Li, Egon Sohn.

**The Hall effect occurs only in systems with broken time-reversal symmetry, such as solids under an external magnetic field in the ordinary Hall effect and magnetic materials in the anomalous Hall effect (AHE) [1]. Here we show a new Hall effect in a nonmagnetic material under zero magnetic field, in which the Hall voltage depends quadratically on the longitudinal current [2-6]. We observe the effect (referred to as nonlinear AHE) in two-dimensional T$_d$-WTe$_2$, a semimetal with broken inversion symmetry and only one mirror line in the crystal plane. Our angle-resolved electrical measurements reveal that the Hall voltage changes sign when the bias current reverses direction; it maximizes (vanishes) when the bias current is perpendicular (parallel) to the mirror line. The observed effect can be understood as an AHE induced by the bias current which generates an out-of-plane magnetization. The temperature dependence of the Hall conductivity further suggests that both intrinsic Berry curvature dipole and extrinsic spin-dependent scatterings contribute to the observed nonlinear AHE. Our results open the possibility of exploring the intrinsic Berry curvature effect in nonlinear electrical transport in solids [3-7].**

Unlike the linear Hall effect that has to vanish to satisfy the Onsager's reciprocity relation in a time-reversal invariant system, in principle, the nonlinear Hall effect does not have to vanish [8]. On the other hand, a second-order nonlinear effect occurs only in systems with broken inversion symmetry [9]. Atomically thin T$_d$-WTe$_2$ possesses all the right symmetries to realize the second-order nonlinear anomalous Hall effect (AHE) with an in-plane Hall conductivity under an in-plane bias current. Monolayer WTe$_2$ of the T$_d$/T' polytype consists of a layer of W atoms sandwiched between two layers of Te atoms in a distorted octahedral coordination [10] (Fig. 1a). It is centrosymmetric with one mirror line (dashed line, Fig. 1a) along the crystal b-axis. Multilayer T$_d$-WTe$_2$ is formed by stacking monolayers with rotated alternating layers by 180 degrees [10] (Fig. 1a). It is non-centrosymmetric and has point group $Pm$ (Ref. [11]). In contrast to the bulk (point group $Pmn2_1$) [10], the screw-axis and glide plane symmetries are broken at the surfaces to allow an in-plane polar axis along the mirror line. Pristine T$_d$-WTe$_2$ is a semimetal with nearly compensated electron and hole densities down to a thickness of three layers [12-15]. An array of quantum revelations has been recently reported in this system including a two-dimensional (2D) topological insulator in the monolayer limit [16-19], superconductivity induced by electrostatic doping in monolayers [20, 21], and a switchable ferroelectric metal in two- and three-layers [22]. Here we investigate the nonlinear electrical properties of atomically thin $T_d$-WTe$_2$.

In our experiment, $T_d$-WTe$_2$ samples with a thickness of 4 - 8 layers have been studied. They were fabricated by mechanical exfoliation from bulk crystals (HQ Graphene) and were capped by hexagonal boron nitride thin layers to prevent sample degradation. The sample thickness was determined by the atomic force microscopy. The crystal orientation was inferred from the polarized Raman spectroscopy [11] (see Supplementary Sect. 1 for details). Two types of devices were employed (Fig. 1b). The first type is of the Hall bar geometry with the Pt electrodes aligned with the principal crystal axes a and b. The second type is of the circular disk geometry with 16 Pt electrodes also aligned with respect to the crystal axes. The latter was essential for angle-resolved electrical measurements. (See Methods for details on the sample and device fabrication.)

Figure 1c and 1d show the result of basic electrical characterization of a typical Hall bar device. The temperature dependence of the longitudinal resistivity $\rho_\parallel$ along the a-axis (Fig. 1c) shows a typical metallic behavior below ~ 100 K with a residual resistivity of ~ 50 μΩ cm. Above 100 K, the dependence deviates from being linear, which is likely a manifestation of a non-degenerate electron gas in semi-metallic WTe$_2$ (Ref. [15]). Because of its low carrier density or small Fermi energy especially in thin samples with reduced carrier density, the electron or hole gas can be thermally excited easily. Figure 1d is the longitudinal $\rho_\parallel$ and Hall resistivity $\rho_\perp$ as a function of out-of-plane magnetic field at 1.8 K. The large non-saturating magnetoresistance up to 14 T indicates a nearly compensated electron and hole density [12, 14, 15]. The Hall resistivity depends approximately linearly on the magnetic field. Furthermore, Landau levels start to form around 7 T (plateaus in the Hall resistivity), indicative of high sample quality. Using the two-carrier analysis of the magnetic field dependence of $\rho_\parallel$ and $\rho_\perp$ (Ref. [14, 15, 23]), we extract the carrier densities ~$1.26 \times 10^{13}$ cm$^{-2}$ (electron) and ~$1.16 \times 10^{13}$ cm$^{-2}$ (hole) and the mobilties ~1580 cm$^2$V$^{-1}$s$^{-1}$ (electron) and ~960 cm$^2$V$^{-1}$s$^{-1}$ (hole). The mobilities along the b-axis are typically 2 - 3 times smaller. The behavior of other Hall bar devices (summarized in Supplementary Sect. 3) is similar. These results are consistent with the literature [12, 14, 15, 23].

To measure nonlinear transport in few-layer WTe$_2$, we bias the devices with an ac current at a fixed frequency (137 Hz) and record the longitudinal and transverse voltage drops at both the fundamental and second-harmonic frequencies. (See Methods for details on the electrical measurements.) Figure 2 is the result from a 5-layer Hall bar device at 1.8 K under *zero* magnetic field. At the fundamental frequency (Fig. 2a), the longitudinal voltage $V_\parallel$ increases linearly with current $I$ till sample heating becomes significant at large bias currents, whereas the transverse voltage $V_\perp$ remains zero. The linear dependence indicates the formation of ohmic contacts of few-layer WTe$_2$ to Pt electrodes. The absence of a linear transverse voltage is expected in few-layer WTe$_2$ (a nonmagnetic semimetal) under zero magnetic field. It shows that longitudinal-transverse coupling is negligible in our Hall bar device. On the other hand, both second-harmonic longitudinal and transverse voltages are nonzero. The former has a significant contribution from sample heating under large bias currents. We focus on the background free transverse response $V_\perp^{2\omega}$ in Fig. 2b. The response is on the order of ~ 0.1% of $V_\parallel$. It scales linearly with square of the current or, equivalently, $V_\parallel$. In addition, it switches sign when the current direction reverses, which also excludes any contribution from sample heating. The observed nonlinear transverse voltage is thus a second-order nonlinear response to the bias current. Below we use the linear slope of $V_\perp^{2\omega}$ vs. $(V_\parallel)^2$ to quantify the effect.

Next we investigate the nonlinear transverse response as a function of direction of current injection with respect to the crystal orientation. Devices of the disk geometry (Fig. 1b) were utilized. The current $I$ was injected through one of the sixteen electrodes, and the voltage drop across the electrodes along the diameter and perpendicular to the diameter were measured. The direction of the current injection is denoted by angle $\theta$, which is measured from the mirror line (*i.e.* crystal b-axis) as shown in Fig. 1a. Measurements at sixteen angles have been obtained in Fig. 3 from an 8-layer device. Consistent results were observed in all disk devices studied in this work (See supplementary Sect. 4 for details). The first-harmonic longitudinal and transverse voltages are linear with current and the slopes defined as the longitudinal and transverse resistances, $R_\parallel (\equiv \frac{V_\parallel}{I})$ and $R_\perp (\equiv \frac{V_\perp}{I})$, respectively, are presented in Fig. 3a. The second-harmonic transverse response is quadratic with current (or, equivalently, $V_\parallel$) for all angles (Fig. 3b). The slope of $V_\perp^{2\omega}$ vs. $(V_\parallel)^2$ as a function of $\theta$ is summarized in Fig. 3c. The longitudinal and transverse resistances both show a two-fold angular dependence. This is consistent with the crystal symmetry of few-layer $T_d$-WTe$_2$ (Ref. [11]). The observed angular dependences can be expressed by considering the resistance tensor in a rotated reference frame as $R_\parallel(\theta) = R_b \cos^2\theta + R_a \sin^2\theta$ and $R_\perp(\theta) = (R_b - R_a)\sin\theta\cos\theta$, where $R_a$ and $R_b (> R_a)$ are the resistance along the crystal a- and b-axis, respectively. The fit (dashed lines, Fig. 3a) yields a resistance anisotropy $r(\equiv \frac{R_a}{R_b})$ of about 0.3. This value is nearly temperature independent, particularly, for temperatures below 100 K (Supplementary Sect. Fig. S2).

In contrast, the nonlinear response shows a one-fold angular dependence. The maximum response occurs when the current is injected perpendicular to the mirror line (*i.e.* along the a-axis), and vanishes when the current is injected parallel to the mirror line (*i.e.* along the b-axis). The response switches sign when the bias current reverses direction. We can express the nonlinear response through the second-order nonlinear susceptibility tensor $\chi_{ijk}^{(2)}$ for the $Pm$ point group [9] (see Supplementary Sect. 5 for derivation)

$$V_\perp^{2\omega}/(V_\parallel)^2 \propto \sin\theta \frac{d_{12}r^2\sin^2\theta + (d_{11} - 2d_{26}r^2)\cos^2\theta}{(\cos^2\theta + r\sin^2\theta)^2}, \qquad (1)$$

where $d_{ij}$'s are the non-vanishing elements of $\chi_{ijk}^{(2)}$. The overall $\sin\theta$ factor dictates that the effect is present only when there is a current component perpendicular to the mirror line. Equation (1) with three free parameters, including an overall amplitude, $d_{12}$, and $d_{11} - 2d_{26}r^2$, captures the experimental angular dependence of Fig. 3c very well (dashed line). The observed effect is thus fully consistent with the interpretation of a second-order nonlinear response to bias current based on the symmetry analysis of few-layer $T_d$-$WTe_2$ crystals.

Finally, we examine the microscopic mechanisms of the observed nonlinear effect by studying its dependence on the carrier scattering processes. We vary the material's conductivity (or resistivity) by changing the sample temperature (Fig. 4a). We focus on the Hall bar devices, in which the few-layer $T_d$-$WTe_2$ crystals are aligned for maximum nonlinear response. The Hall bar geometry also allows us to obtain the electric fields (and conductivities) from the voltage drops through the device dimensions, $E_\perp^{2\omega} \equiv \frac{V_\perp^{2\omega}}{L_\perp}$ and $E_\parallel \equiv \frac{V_\parallel}{L_\parallel}$, where $L_\parallel \approx 6$ μm and $L_\perp \approx 9.2$ μm are the longitudinal and transverse length of the Hall bar device. Figure 4b shows that similar to what has been observed at 1.8 K, the transverse electric field $E_\perp^{2\omega}$ depends linearly on $(E_\parallel)^2$ at all temperatures, and changes sign when current or $E_\parallel$ reverses direction. The slope, however, decreases monotonically with increasing temperature. In Fig. 4c, we extract the slope for temperatures below 100 K, at which the electron gas in $T_d$-$WTe_2$ remains largely degenerate. Interestingly, the dependence resembles that of the longitudinal conductivity $\sigma$ (along the a-axis) shown in Fig. 4a, where $\sigma$ increases with decreasing temperature and saturates to a value limited by disorder scattering at low temperatures. A more careful analysis in Fig. 4d shows that $\frac{E_\perp^{2\omega}}{(E_\parallel)^2}$ scales linearly with $\sigma^2$ (dashed line):

$$\frac{E_\perp^{2\omega}}{(E_\parallel)^2} = \xi\sigma^2 + \eta, \qquad (2)$$

with constant $\xi$ and $\eta$ representing the slope and intersect, respectively.

The scaling result of Eqn. (2) inferred from our experiment sheds light on the mechanisms that give rise to the second-order nonlinear response to current in few-layer $T_d$-$WTe_2$. We rewrite the left hand ride of Eqn. (2) in terms of the ratio of the conductivities instead of the electric fields: $\frac{1}{E_\parallel}\left(\frac{E_\perp^{2\omega}}{E_\parallel}\right) = \frac{1}{E_\parallel}\left(\frac{\sigma_{AH}}{r\sigma}\right)$, where $\sigma_{AH}$ denotes the off-diagonal element of the conductivity tensor. Since the longitudinal conductivity $\sigma$ depends linearly on scattering time $\tau$ along the a-axis, Eqn. (2) states that $\sigma_{AH}$ has two contributions, which scale as $\tau$ and $\tau^3$, respectively. (Note that the resistance anisotropy $r$ is practically temperature independent within this temperature range as shown in Fig. S2.) This excludes effects such as ratchet effect that arises from an inversion asymmetric scattering time and has been shown to scale as $\tau^2$ (Ref. [3, 24]).

Our results are fully consistent with a current-induced AHE [3-6], in which an in-plane current generates an out-of-plane magnetization that acts as the spontaneous magnetization in magnetic materials in the AHE (Fig. 1a). And only the current component that is perpendicular to the polar axis (crystal b-axis) can generate the magnetization [25]. Such current-induced magnetization has been observed in uniaxially strained monolayer MoS$_2$ (which has the same symmetry as few-layer T$_d$-WTe$_2$) using optical Kerr microscopy [26, 27] and has been demonstrated as a spin source at the WTe$_2$/ferromagnet interfaces based on spin-orbit torque [28, 29]. The current-induced magnetization has been shown to scale linearly with the net electronic momentum gained under an in-plane bias that is perpendicular to the mirror line, $M \propto eE_\parallel \tau \sin\theta$ ($e$ denoting the elementary charge) [25]. The magnetization thus scales as $\tau^1$. On the other hand, it is known that the AHE in magnetic materials can arise from both intrinsic (Berry curvature) and extrinsic (spin dependent scattering) effects [1]. When sample temperature is varied, the anomalous Hall conductivity in magnetic materials has been shown to scale as $\tau^2$ for the skew scattering contribution [30, 31], and as $\tau^0$ for the intrinsic and side-jump contributions [1]. Combining with the linear dependence on $\tau$ of the current-induced magnetization, these effects fully account for the two terms in Eqn. (2). In Fig. 4c the intersect $\eta$ is only slightly over 10% of the maximum value, which indicates that the skew scattering contribution dominates the observed nonlinear AHE in this device. In other devices (Supplementary Sect. 3), however, the values of $\xi$ and $\eta$ vary due to changes in the doping density, mobility and sample thickness. The relative importance between the intrinsic and the extrinsic contributions is therefore sample dependent. To provide an order of magnitude estimate to the Berry curvature dipole from the intersect $\eta \approx (0.15 - 1) \times 10^{-3}$ μmV$^{-1}$, we neglect the side-jump contribution. Following the formalism developed by Sodemann and Fu [3], we estimate the Berry curvature dipole to be $\eta * \epsilon_F/e \sim 0.1 - 0.7$ Å (see Methods), where the Fermi energy $\epsilon_F \sim 50$ meV is obtained from *ab initio* band structure calculations of bulk crystals [12, 32] (note that our atomically thin samples likely have a smaller Fermi energy due to quantum confinement [16]). This value for the Berry curvature dipole is on par with the value predicted for strained monolayer semiconductor transition metal dichalcogenides such as MoS$_2$ under 10 – 50% uniaxial strain [3, 6].

In conclusion, we have demonstrated a new nonlinear Hall effect in few-layer T$_d$-WTe$_2$ under zero magnetic field through angle-resolved electrical measurements. The study indicates that both intrinsic Berry curvature dipole and extrinsic spin-dependent scatterings contribute to the nonlinear Hall conductance. Although future theoretical studies including spin dependent scattering contributions are needed for a comprehensive description of the effect, the nonlinear AHE observed here opens up a new avenue for exploring the Berry curvature dipole physics in nonmagnetic solids and for spin-charge conversion applications.

**Methods**
**Sample and device fabrication**
Van der Waals heterostructures of few-layer WTe$_2$ and hexagonal boron nitride (hBN) were fabricated by the layer-by-layer dry transfer method [33]. In this process, few-layer WTe$_2$ and hBN were exfoliated from high quality bulk crystals onto silicon substrates in a nitrogen-filled glovebox. The hBN and WTe$_2$ flakes were picked up by a stamp consisting of a thin film of polycarbonate (PC) on polydimethylsiloxane (PDMS). The stack was then released onto a Si substrate with pre-patterned Pt electrodes. The residual PC was dissolved in chloroform and acetone. The thickness of WTe$_2$ layers was first estimated by optical contrast, and then measured by atomic force microscopy (AFM) after device fabrication. To avoid geometrical complications on the current path, rectangular strips were chosen for the Hall bar devices. The crystal a-axis is aligned parallel to the electrodes for current injection. For the disk devices, the WTe$_2$/hBN stacks were patterned into circular shape using a mask that is made of e-beam exposed PMMA A4/PMMA M2 bilayers followed by reactive ion etching using SF$_6$ gas. The devices were then left in acetone for 30 minutes to remove the PMMA residues.

**Linear and nonlinear transport measurements**

The transport measurements were carried out in a Quantum Design Physical Property Measurement System (PPMS) and an Oxford Teslatron system down to 1.8 K. An AC current at 137 Hz was applied to the devices. The longitudinal and transverse voltage drops at both the fundamental and second-harmonic frequencies were measured by lock-in amplifiers. The longitudinal resistivity of the Hall bar devices was obtained as $\rho_\parallel = \frac{V_\parallel}{I} \cdot \frac{L_\perp \cdot d}{L_\parallel}$, where $I$ is the current amplitude, $V_\parallel$ is the first-harmonic longitudinal voltage, $L_\parallel$ and $L_\perp$ are the longitudinal and transverse dimensions of the Hall bar device, respectively, and $d$ is the sample thickness. Similarly, the transverse resistivity $\rho_\perp$ was obtained by using the first-harmonic transverse voltage $V_\perp$ in place of $V_\parallel$. The longitudinal conductivity is related to the resistivity as $\sigma = 1/\rho_\parallel$.

To extract the angular dependence of the nonlinear anomalous Hall effect in disk devices, the second-harmonic transverse voltage $V_\perp^{2\omega}$ was first anti-symmetrized between two opposing current directions, i.e. $[V_\perp^{2\omega}(\theta) - V_\perp^{2\omega}(\theta + 180°)]/2$, where $\theta$ is the angle for current injection. At each angle, the linear slope was determined from the dependence of the anti-symmetrized $V_\perp^{2\omega}$ as a function of $(V_\parallel)^2$. Anti-symmetrization is needed here because of the longitudinal-transverse coupling at current injection directions not along the crystal axes. No anti-symmetrization is needed along the high-symmetry axes.

**Estimate of the Berry curvature dipole**

In the low-frequency limit, as in our experiment, the intrinsic contribution to the nonlinear Hall conductivity $\sigma_{AH} \approx \frac{\pi}{2} \frac{kDG_0}{d}$ can be expressed through the Berry curvature dipole $D$ (Ref. [3]). In two dimensions, it has the dimension of length. Here $G_0$ is the conductance quantum and $\hbar k \equiv eE_\parallel \tau$ is the net electronic momentum gained under in-plane bias $E_\parallel$ ($\hbar$, $e$ and $\tau$ denote the Planck constant, elementary charge, and carrier scattering time along the a-axis for our device geometry, respectively). On the other hand, the longitudinal conductivity can be expressed as $\sigma = G_0 v_F k_F \tau/2d$, where $v_F$ and $k_F$ are the Fermi velocity and Fermi vector and are related to the Fermi energy as $\epsilon_F \sim \hbar v_F k_F$. We thus obtain $\frac{\sigma_{AH}}{\sigma E_\parallel} \approx \frac{\pi D}{(\hbar v_F k_F)/e} \sim \frac{\pi D}{\epsilon_F/e}$. This was used to evaluate the Berry curvature dipole $D$ from the value of $\frac{\sigma_{AH}}{\sigma E_\parallel}$ in the limit of $\sigma \to 0$ when the intrinsic effect dominates the nonlinear Hall conductivity.

**Acknowledgements**
We thank Gregory Stiehl for fruitful discussions on the crystal symmetry properties of multilayer $T_d$-WTe$_2$.

## Figures

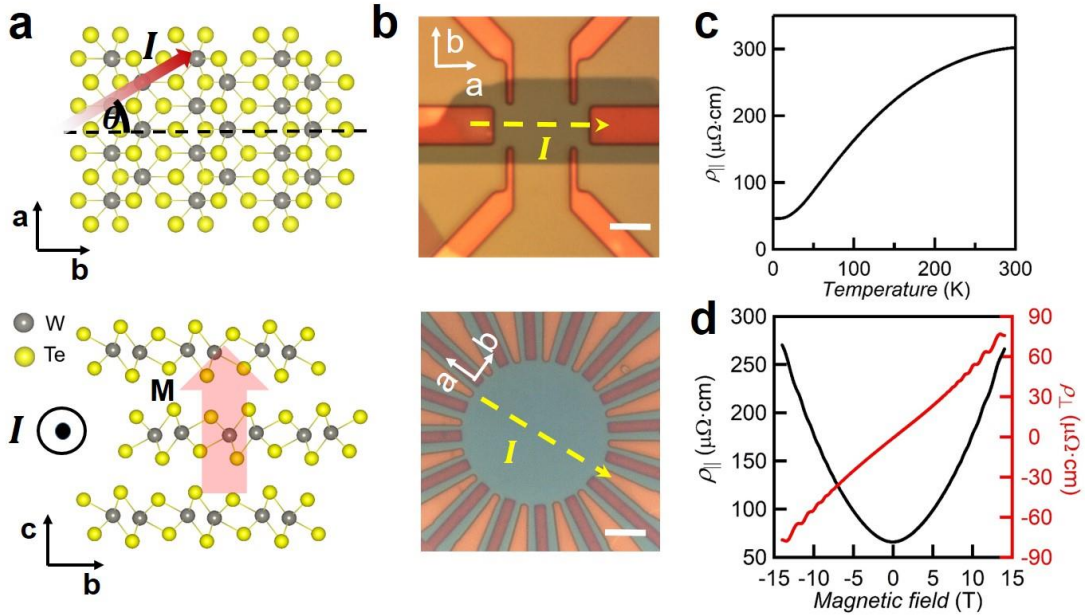

**Figure 1 | Atomic structure and basic characterization of few-layer WTe$_2$. a,** a-b plane of monolayer and b-c plane of few-layer T$_d$-WTe$_2$. W atoms are in grey and Te atoms in yellow. When an in-plane current $I$ is injected at angle $\theta$ ($\neq 0$) from the mirror line (dashed line), an out-of-plane magnetization (**M**) is generated. **b,** Optical image of a typical Hall bar (top) and disk (bottom) devices used in this study. The scale bars are 5 $\mu$m. The crystal orientation (a- and b-axis) were determined by the polarized Raman spectroscopy. WTe$_2$ samples are in grey, Pt electrodes in orange, and Si substrates are in yellow-orange for the Hall bar device and red-orange for the disk device, respectively. Dashed yellow lines indicate the current injection direction. **c, d,** Temperature dependence of the longitudinal resistivity (**c**) and magnetic-field dependence of the longitudinal and transverse resistivity (**d**) of a Hall bar device at 1.8 K.

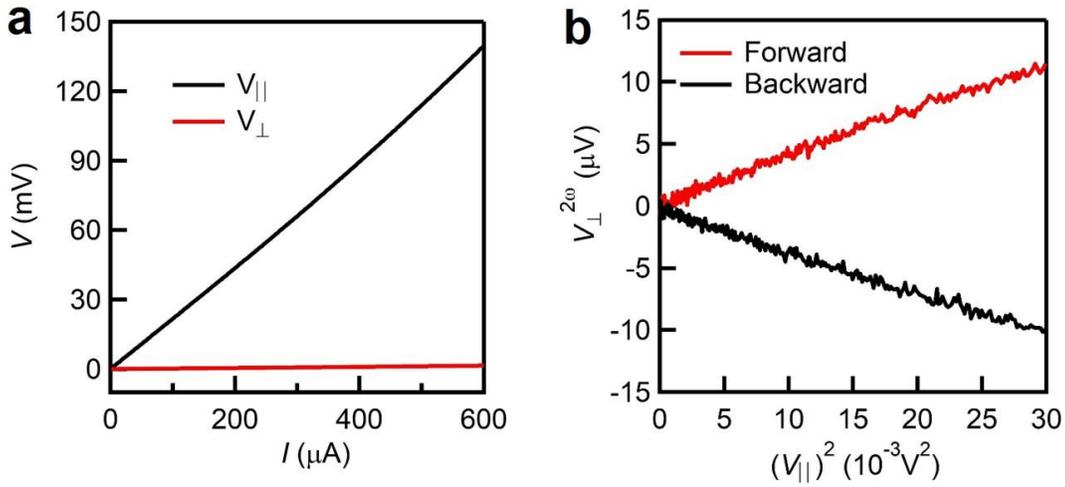

**Figure 2 | Nonlinear anomalous Hall effect (AHE). a,** Dependence of first-harmonic longitudinal and transverse voltages as a function of ac current amplitude in a 5-layer WTe$_2$ Hall bar device at 1.8 K under zero magnetic field. Current is injected along the crystal a-axis. **b,** The second-harmonic transverse voltage depends linearly on the square of the first-harmonic longitudinal voltage. It changes sign when the current direction reverses.

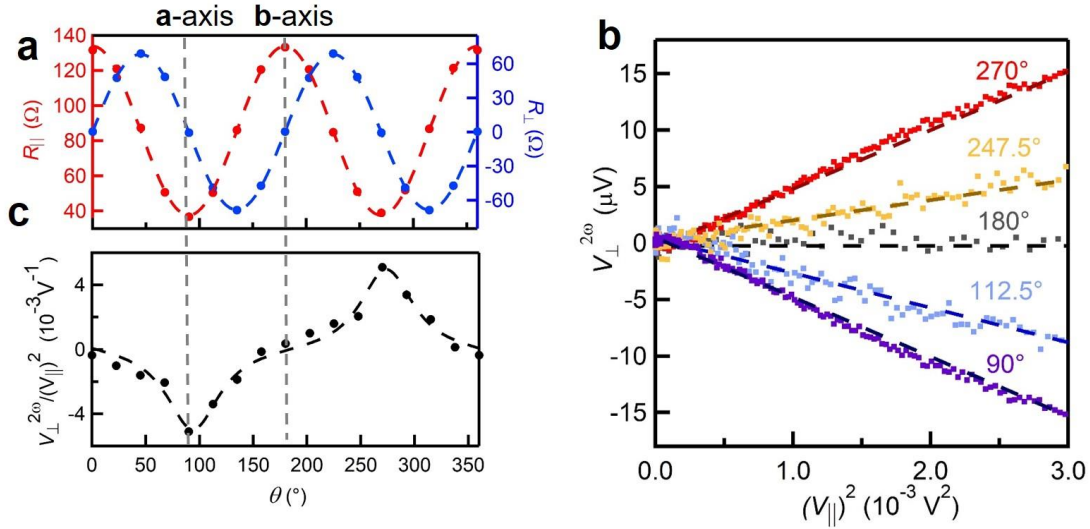

**Figure 3 | Angular dependence of the nonlinear AHE. a,** Longitudinal and transverse resistance of a disk device as a function of direction of current injection. **b,** The second-harmonic transverse voltage depends linearly on the square of the first-harmonic longitudinal voltage for different current injection angles. **c,** The nonlinear Hall effect (slope of the dependences in **b**) as a function of direction of current injection. All measurements were performed at 1.8 K. The symbols in **a** and **c** are experimental data and the dashed lines are fits to the model described in the text. The two vertical lines indicate the direction of the crystal a- and b-axis.

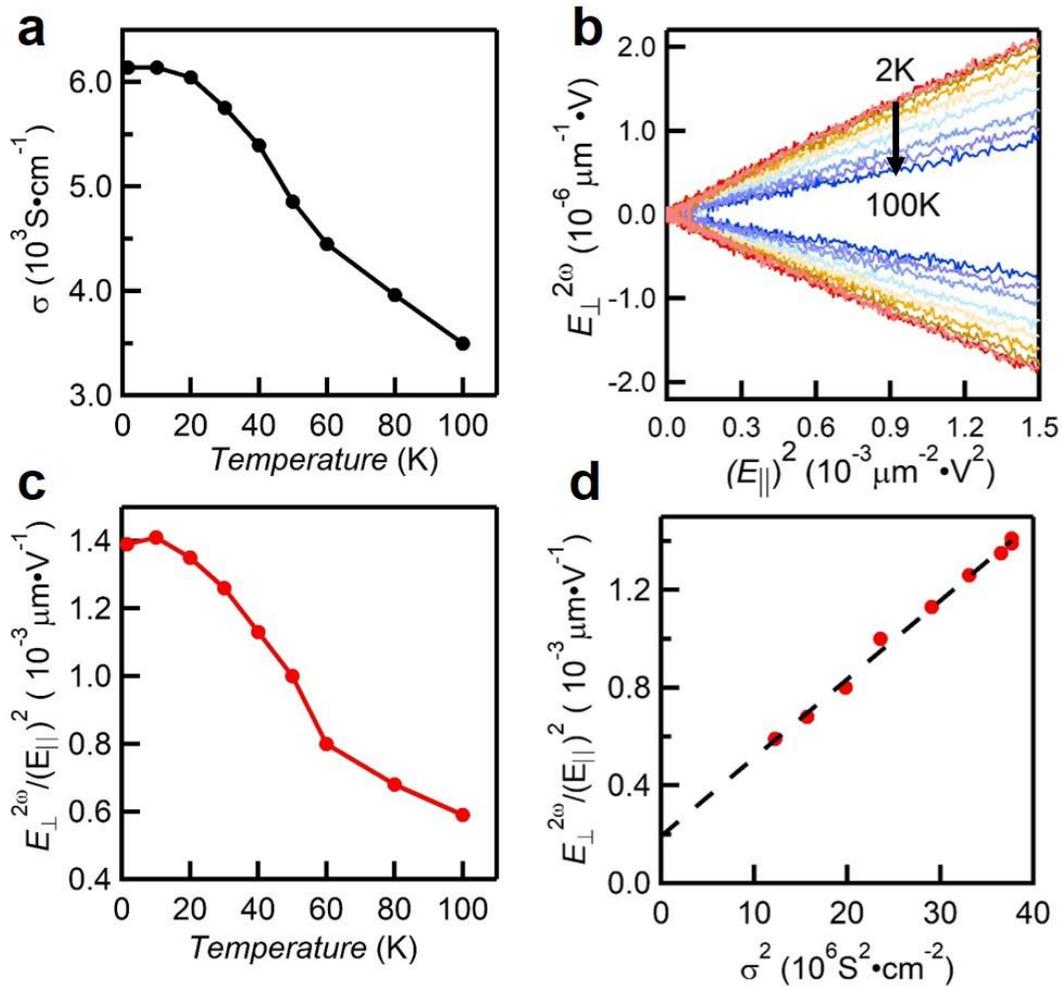

**Figure 4 | Temperature dependence of the nonlinear AHE. a,** Longitudinal conductivity as a function of temperature of the Hall bar device employed in Fig. 2. **b,** The second-harmonic transverse field depends linearly on the square of the first-harmonic longitudinal field at temperature ranging from 2 to 100 K. It changes sign when the current direction reverses. **c, d,** The nonlinear Hall effect (slope of the dependences of **b**) as a function of temperature (**c**) and square of the longitudinal conductivity (**d**). The lines in **a** and **c** are guides to the eye. The dashed line in **d** is a linear fit to the experimental data (symbols).

# Supplementary information for
# Observation of the nonlinear anomalous Hall effect in 2D WTe$_2$

Kaifei Kang, Tingxin Li, Egon Sohn, Jie Shan, and Kin Fai Mak

## 1. Raman spectroscopy of few-layer T$_d$-WTe$_2$

Raman spectroscopy in the parallel polarization configuration was used to determine the crystal orientation of few-layer T$_d$-WTe$_2$ following Ref. [1]. A linearly polarized HeNe laser beam (633 nm) was used to excite samples under normal incidence. It impinged the sample's a-b plane after passing through a half waveplate, which varies the polarization direction with respect to the crystal axes. The scattered light passed the same waveplate and then a linear polarizer, which selects the component parallel to the incident beam polarization, and was recorded by a spectrometer equipped with a charge-coupled device (CCD). Figure S1a shows the Raman spectrum for three representative polarization directions. Five Raman peaks can be identified. They correspond to the A1 modes of T$_d$-WTe$_2$. The intensity of each mode depends strongly on the polarization. Figure S1b is the contour plot of the spectrum showing the two high-energy modes (at ~ 160 cm$^{-1}$ and ~ 210 cm$^{-1}$, respectively) as a function of the polarization angle. A two-fold symmetry is observed, which is consistent with the crystal symmetry. These two modes are out of phase. In particular, the polarization direction corresponding to the maximum of the 160 cm$^{-1}$ mode and the minimum of the 240 cm$^{-1}$ mode is along the b-axis of the crystal.

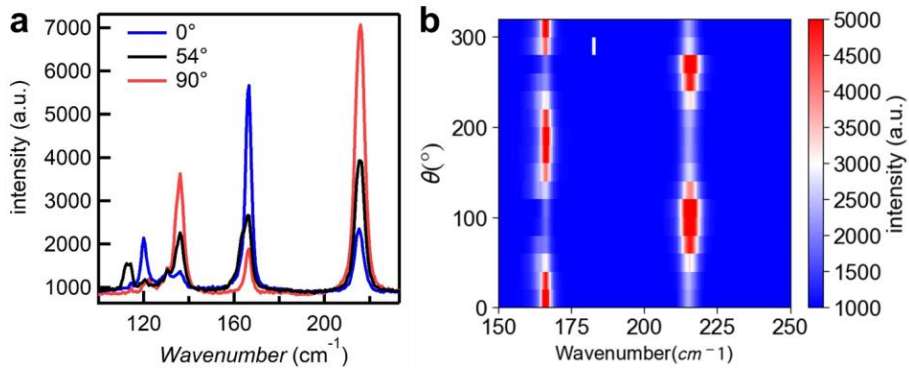

**Fig. S1 | Raman spectroscopy of few-layer T$_d$-WTe$_2$ in the parallel polarization configuration. a,** Raman spectrum for three selected polarization angles. **b,** Contour plot of the Raman spectrum (showing modes at ~ 160 and 210 cm$^{-1}$) as a function of polarization angle.

## 2. Resistance anisotropy

In the main text we show a two-fold symmetry as a function of current injection direction for both the longitudinal and transverse resistance in a disk device at 1.8 K. Figure S2a shows the angular dependence of the parallel resistance at varying temperatures up to 200 K. Our result shows that the overall resistance increases with increasing temperature. The two-fold symmetry is observed at all temperatures with the resistance anisotropy ratio weakly dependent on temperature (Fig. S2b). It remains ~ 3 up to 150 K.

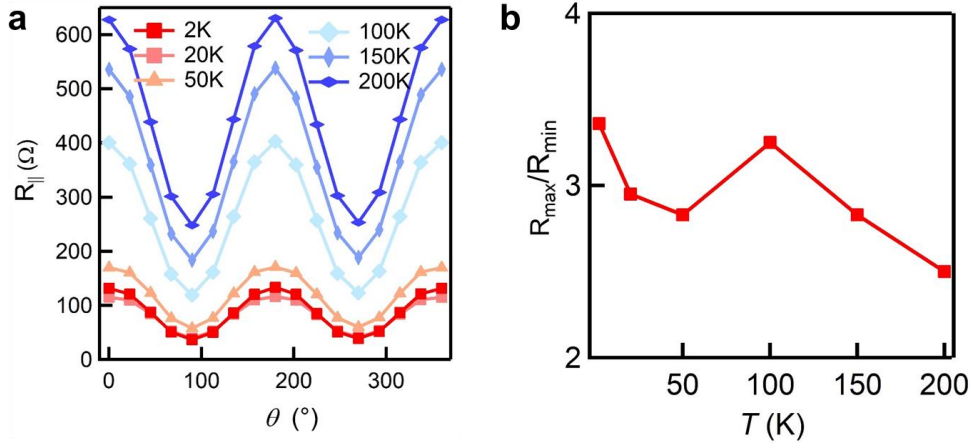

**Fig. S2 | Resistance anisotropy in few-layer T$_d$-WTe$_2$. a,** Longitudinal resistance as a function of angle of current injection at varying temperatures. **b,** Resistance anisotropy weakly depends on temperature.

## 3. Additional Hall bar devices

In the main text, we have shown the result from Hall bar device S18. Below we show the results from two other Hall bar devices (S13 and S19), which are of the same longitudinal and transverse dimensions and similar sample thickness. In all Hall bar devices the crystal a-axis is aligned with the electrodes for current injection. Figure S3a-c are the optical images of the devices. Hall measurements were carried out to extract the density and mobility of the electrons and holes. Figure S3d-f show the magnetic field ($B$) dependence of the longitudinal ($\rho_\parallel$) and transverse ($\rho_\perp$) resistivity of the devices at 1.8 K. A semi-classical two-carrier model was employed to describe these dependences [2,3]:

$$\rho_\parallel = \frac{1}{e}\frac{n\mu_n + p\mu_p + (n\mu_p + p\mu_n)\mu_n\mu_p B^2}{(n\mu_n + p\mu_p)^2 + (n-p)^2\mu_n^2\mu_p^2 B^2},$$

$$\rho_\perp = \frac{1}{e}\frac{(p\mu_p^2 - n\mu_n^2)B + (p-n)\mu_n^2\mu_p^2 B^3}{(n\mu_n + p\mu_p)^2 + (n-p)^2\mu_n^2\mu_p^2 B^2}.$$

(1)

Here $n$, $p$, $\mu_n$ and $\mu_p$ denote the electron density, hole density, electron mobility, and hole mobility, respectively. The fitting results to Eqn (1) are summarized in Table S1.

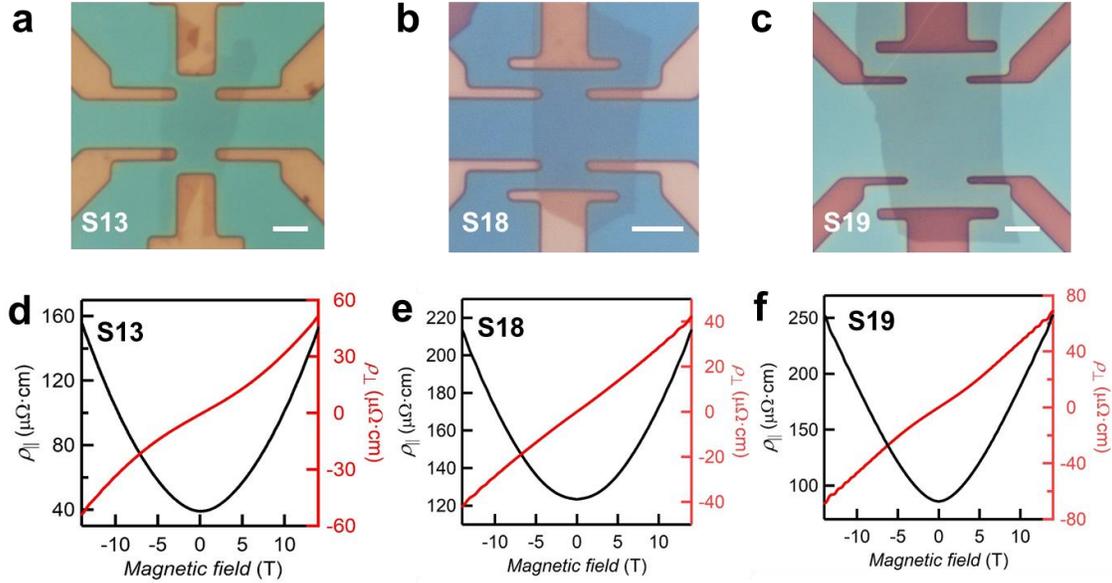

**Fig. S3 | Linear transport properties.** Optical image of Hall bar device S13 (**a**), S18 (**b**), and S19 (**c**). Scale bars are 5 $\mu m$. Magnetic-field dependence of the longitudinal and transverse resistivity for device S13,(**d**), S18 (**e**), and S19 (**f**).

**Table S1.** Density and mobility of carriers at 1.8 K.

| Sample# | Thickness | RRR | $\rho(T=300K)$ $(\mu\Omega \cdot nm)$ | $n$ $(10^{13}/cm^2)$ | $p$ $(10^{13}/cm^2)$ | $\mu_e$ $cm^2/(V \cdot S)$ | $\mu_h$ $cm^2/(V \cdot S)$ |
|---|---|---|---|---|---|---|---|
| 13 | 5L, 3.5nm | 6.55 | 303.8 | 2.44±0.02 | 1.86±0.02 | 1580±16 | 1000±10 |
| 18 | 6L, 4.2nm | 2.91 | 312.5 | 2.44±0.02 | 2.18±0.02 | 710±7 | 533±5 |
| 19 | 5L, 3.5nm | 7.72 | 527.1 | 1.67±0.02 | 1.49±0.02 | 1166±12 | 793±8 |

Figure S4 shows that the second-harmonic transverse voltage in all three devices scales linearly with the square of the first-harmonic longitudinal voltage under zero magnetic field. For Device S18 we also show that the second-harmonic transverse voltage changes sign when the current direction reverses (blue line, Fig. S4b).

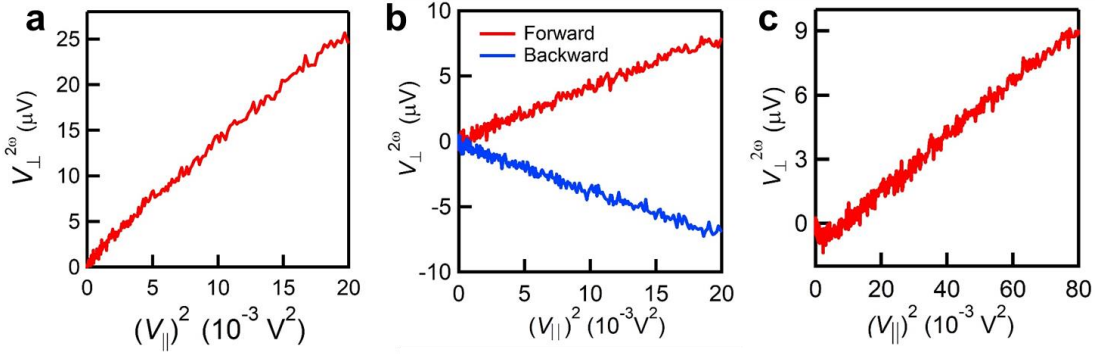

**Fig. S4 | Nonlinear anomalous Hall effect (AHE).** Second-harmonic transverse voltage $V_\perp^{2\omega}$ as a function of $(V_\parallel)^2$ for device S13 (**a**), S18 (**b**) and S19 (**c**) at 1.8K.

We investigate the mechanism of the nonlinear anomalous Hall effect (AHE) by varying the sample temperature. We have shown in Fig. 4d of the main text the linear dependence of $\frac{E_\perp^{2\omega}}{(E_\parallel)^2}$ on $\sigma^2$ up to 100 K. Here $E_\parallel$ and $E_\perp^{2\omega}$ are the longitudinal and second-order transverse electric field, respectively, and $\sigma$ is the longitudinal conductivity. In Fig. S5 we show the result of all three devices from 2 K to 100 K. It is clear that $\frac{E_\perp^{2\omega}}{(E_\parallel)^2} = \xi\sigma^2 + \eta$ describes all three devices satisfactorily with $\xi$ and $\eta$ as fitting parameters (summarized in Table S2). These values vary by a few times among the different samples. This is not surprising considering the differences in the carrier mobility and doping densities (Table S1).

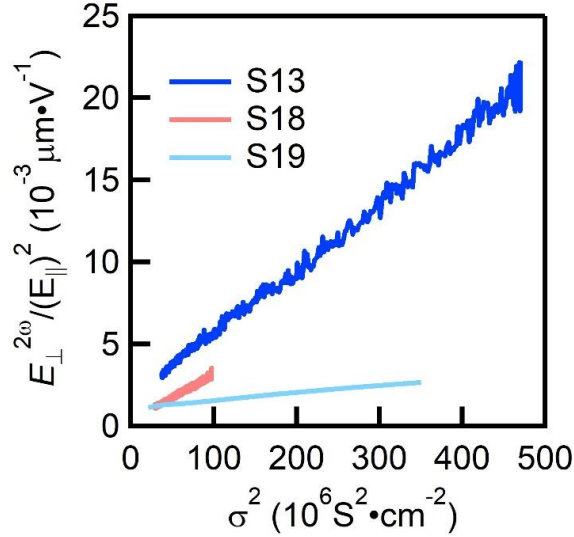

**Fig. S5 |** Nonlinear AHE can be described by $\frac{E_\perp^{2\omega}}{(E_\parallel)^2} = \xi\sigma^2 + \eta$ in all three devices from 2 K to 100 K.

**Table S2.** Fitting parameters for $\xi$ and $\eta$ ($\frac{E_\perp^{2\omega}}{(E_\parallel)^2} = \xi\sigma^2 + \eta$).

| Device# | $\sigma(T \to 0)$ $(10^3 S \cdot cm^{-1})$ | $\xi$ $(10^{-19} m^3 \cdot V^{-1} S^{-2})$ | $\eta$ $(10^{-3} \mu m \cdot V^{-1})$ |
|---|---|---|---|
| 13 | 19.0 | 0.030±0.001 | 0.99±0.01 |
| 18 | 6.4 | 0.029±0.002 | 0.147±0.001 |
| 19 | 12.7 | 0.006±0.001 | 0.56±0.02 |

## 4. Additional disk devices

In the main text, we have shown the result from one disk device (S1). Below we show the results from two other disk devices (S2 and S3), which are of the same dimension and similar sample thickness. Figure S6a-c show the optical image of all three disk devices. Figure S6d-f show that in all devices the second-harmonic transverse voltage scales quadratically with the longitudinal voltage; it switches sign when the current direction reverses; and for certain current injection angles, the effect vanishes. Figure S6g-i summarize the angular dependence of the effect (symbols) at different temperatures. All of them are all well described by Eqn. (1) of the main text (solid lines).

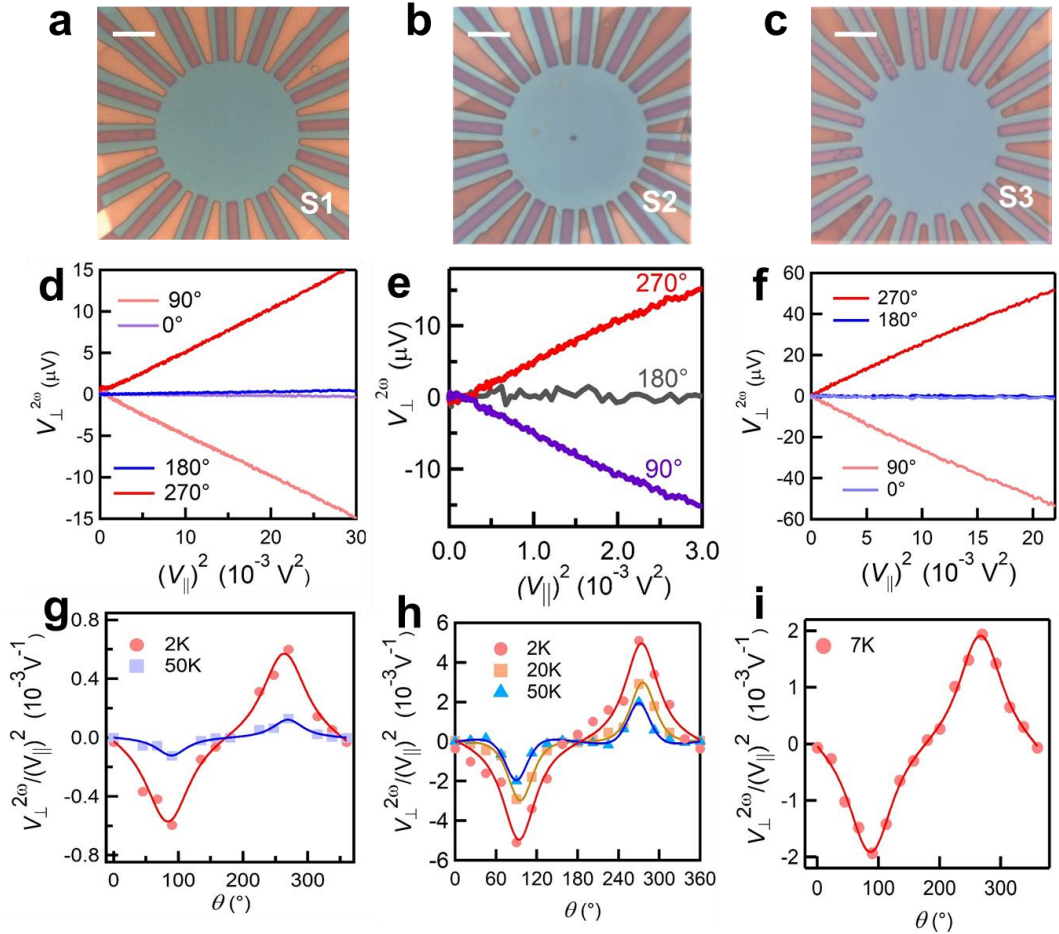

**Fig. S6 | Nonlinear AHE in disk devices. a, - c,** Optical images for disk device S1 (**a**), S2 (**b**) and S3 (**c**). Few-layer WTe$_2$ in these devices was etched into a circular disk shape afterwards. The scale bars in all images are 5um. **d, - f,** $V_\perp^{2\omega}$ vs. $(V_\parallel)^2$ for current injected along different directions; **g, - i,** Angular dependence of the nonlinear AHE in three different devices.

## 5. Symmetry analysis of the second-order nonlinear effect

In general, the second-order nonlinear current density $\vec{j}^{(2)}$ in response to an electric field $\vec{E}$ in a material can be expressed through the material's second-order nonlinear susceptibility $\overleftrightarrow{\chi}^{(2)}$ as $\vec{j}^{(2)} = \overleftrightarrow{\chi}^{(2)} \cdot \vec{E} \cdot \vec{E}$ [4, 5]. For few-layer T$_d$-WTe$_2$ with $Pm$ point group symmetry, the nonlinear susceptibility has the following form

$$\overleftrightarrow{\chi}^{(2)} = 2 \begin{pmatrix} d_{11} & d_{12} & d_{13} & 0 & d_{15} & 0 \\ 0 & 0 & 0 & d_{24} & 0 & d_{26} \\ d_{31} & d_{32} & d_{33} & 0 & d_{35} & 0 \end{pmatrix}, \qquad (2)$$

where the coordinate *x*, *y*, and *z* are along the principal axes of the crystal, namely, parallel to the mirror line in the plane (b-axis), perpendicular to the mirror line in the plane (a-axis), and perpendicular to the plane (c-axis), respectively. For an in-plane electric field $\vec{E} = (E_x, E_y, 0)$, the nonlinear current density $\vec{j}^{(2)}$ is given as

$$\vec{j}^{(2)} = \begin{pmatrix} d_{11}E_x^2 + d_{12}E_y^2 \\ 2d_{26}E_xE_y \\ d_{31}E_x^2 + d_{32}E_y^2 \end{pmatrix}. \tag{3}$$

We can obtain the second-order nonlinear electric field from the nonlinear current density by using the Ohm's law: $\vec{E}^{(2)} = \vec{\rho} \cdot \vec{j}^{(2)}$, where $\vec{\rho} = \begin{pmatrix} \rho_b & 0 & 0 \\ 0 & \rho_a & 0 \\ 0 & 0 & \rho_c \end{pmatrix}$ is the resistivity tensor. The second-order nonlinear electric field in the plane is thus computed as

$$\vec{E}^{(2)} = \begin{pmatrix} \rho_b(d_{11}E_x^2 + d_{12}E_y^2) \\ 2\rho_a d_{26}E_xE_y \end{pmatrix}. \tag{4}$$

Now suppose we bias the sample with an in-plane current at angle $\theta$ measured from the mirror line (x-axis). The current density can be expressed as $\vec{j} = j\begin{pmatrix} \cos\theta \\ \sin\theta \end{pmatrix}$ and the electric field from Ohm's law as $\vec{E} = j\begin{pmatrix} \rho_b\cos\theta \\ \rho_a\sin\theta \end{pmatrix}$. The longitudinal component, i.e. the component parallel to $\vec{j}$, is

$$E_\parallel = j(\rho_b\cos^2\theta + \rho_a\sin^2\theta). \tag{5}$$

On the other hand, the transverse component of the second-order electric field, i.e. the in-plane component perpendicular to $\vec{j}$ is

$$E_\perp^{(2)} = j^2\rho_b^3 \sin\theta \left[d_{12}r^2 \sin^2\theta + (d_{11} - 2d_{26}r^2)\cos^2\theta\right], \tag{6}$$

where $r$ is the resistivity anisotropy defined as $r \equiv \frac{\rho_a}{\rho_b}$. Combining (5) and (6) finally we obtain

$$\frac{E_\perp^{(2)}}{(E_\parallel)^2} = \frac{\rho_b \sin\theta [d_{12}r^2 \sin^2\theta + (d_{11} - 2d_{26}r^2)\cos^2\theta]}{(\cos^2\theta + r\sin^2\theta)^2}. \tag{7}$$

The result describes well the experimental angle dependence of the nonlinear AHE shown in Fig. 3c of the main text and in Fig. S6. At $\theta = \frac{\pi}{2}$, the angle with the largest nonlinear Hall response, $\frac{E_\perp^{(2)}}{(E_\parallel)^2} = \rho_b d_{12}$ so we have $\sigma_{AH} = d_{12}E_\parallel$.